\documentclass[twocolumn,final,aps,prl,showpacs,superscriptaddress,amsmath,amssymb,amsfonts,floatfix]{revtex4-1}
\usepackage{graphicx}
\usepackage{multirow}
\usepackage{epsfig}

\def\be{\begin{equation}}
\def\ee{\end{equation}}

\usepackage{url}
\usepackage{color}
\usepackage[colorlinks,breaklinks,bookmarks=true,citecolor=blue,linkcolor=red,urlcolor=blue]{hyperref}
\definecolor{darkred}{rgb}{0.7,0.0,0}

\begin{document}

\title{Topotactic hydrogen in nickelate superconductors and akin infinite-layer oxides $AB$O$_2$}

\author{Liang Si}
\affiliation{Key Laboratory of Magnetic Materials and Devices \& Zhejiang Province Key Laboratory of Magnetic Materials and Application Technology, Ningbo Institute of Materials Technology and Engineering (NIMTE), Chinese Academy of Sciences, Ningbo 315201, China}
\affiliation{Institute for Solid State Physics, Vienna University of Technology, 1040 Vienna, Austria}

\author{Wen Xiao}
\affiliation{Key Laboratory of Magnetic Materials and Devices \& Zhejiang Province Key Laboratory of Magnetic Materials and Application Technology, Ningbo Institute of Materials Technology and Engineering (NIMTE), Chinese Academy of Sciences, Ningbo 315201, China}

\author{Josef Kaufmann}
\affiliation{Institute for Solid State Physics, Vienna University of Technology, 1040 Vienna, Austria}

\author{Jan M.\,Tomczak}
\affiliation{Institute for Solid State Physics, Vienna University of Technology, 1040 Vienna, Austria}

\author{Yi Lu}
\affiliation{Institute for Theoretical Physics, Heidelberg University, Philosophenweg 19, 69120 Heidelberg, Germany}

\author{Zhicheng Zhong}
\email{zhong@nimte.ac.cn}
\affiliation{Key Laboratory of Magnetic Materials and Devices \& Zhejiang Province Key Laboratory of Magnetic Materials and Application Technology, Ningbo Institute of Materials Technology and Engineering (NIMTE), Chinese Academy of Sciences, Ningbo 315201, China}

\author{Karsten Held}
\email{held@ifp.tuwien.ac.at}
\affiliation{Institute for Solid State Physics, Vienna University of Technology, 1040 Vienna, Austria}

\date{\today}

\begin{abstract}
 Superconducting nickelates appear to be difficult to synthesize. Since the  chemical reduction of  $AB$O$_3$ ($A$: rare earth; $B$ transition metal) with CaH$_2$ may result in both,  $AB$O$_2$ and  $AB$O$_2$H, we calculate the
topotactic H binding energy by density functional theory (DFT). We find intercalating H to be energetically favorable for LaNiO$_2$ but not for Sr-doped NdNiO$_2$. This has dramatic consequences for the  electronic structure as determined by DFT+dynamical mean field theory: that of  3$d^9$ LaNiO$_2$ is similar to (doped) cuprates,   3$d^8$ LaNiO$_2$H is a two-orbital Mott insulator. Topotactic H might hence explain why some nickelates are superconducting and others are not.
\end{abstract}

\maketitle
Most recently superconductivity was found in Nd$_{0.8}$Sr$_{0.2}$NiO$_2$ films grown on SrTiO$_3$ \cite{li2019superconductivity}, a seminal work that  opens the door wide to a new age of superconductivity: the nickelate age.
These novel  (Sr-doped) NdNiO$_2$ superconductors are not only isostructural to 
the well known cuprate superconductor CaCuO$_2$ \cite{siegrist1988parent,Balestrino2002,Orgiani2007,DiCastro2015} but also both,  Ni and Cu, are formally 3$d^9$ in the respective parent compound.

Strikingly different to the cuprates \cite{Bednorz1986} and iron pnictides  \cite{Kamihara2006}, reproducing these outstanding results in isoelectronic compositions appears to be quite challenging. 
In a more bulk-like crystal, no superconductivity was reported for Nd$_{0.8}$Sr$_{0.2}$NiO$_2$ \cite{Li2019}, neither when directly  pulsed laser depositing  Nd$_{0.8}$Sr$_{0.2}$NiO$_x$ \cite{Zhou2019}. Also the parent compound, NdNiO$_2$, is not superconducting \cite{li2019superconductivity} but shows  a resistivity upturn toward low temperatures.
Another nickelate,  LaNiO$_2$, is also isostructural and isovalent, but  is a (bad) metal  \cite{Kaneko2009} with neither superconductivity nor antiferromagnetism \cite{Hayward1999}.

An obvious difference between Nd$_{0.8}$Sr$_{0.2}$NiO$_2$ and Nd(La)NiO$_2$ is doping. However, in contrast to the cuprates, there is already a self-doping of the Ni-bands in Nd(La)NiO$_2$ because one Nd(La) band crosses the Fermi energy \cite{Zhang2019,Nomura2019,Motoaki2019,jiang2019electronic}, hardly hybridizing with the Ni 3$d_{x^2-y^2}$-bands.
Why are some of these nickelates, all of which have a similar paramagnetic (spin-unpolarized) DFT electronic structure \cite{Lee2004,Botana2019,Hirofumi2019,Motoaki2019,hu2019twoband,Wu2019,Nomura2019,Zhang2019,Gao2019,Jiang2019,Liu2019,Werner2020,Geisler2020,Bernardini2020,Choi2020}\footnote{One difference is that Nd has a 4$f$ moment and La has not. But because of the weak hybridization with the Ni-3$d_{x^2-y^2}$, one might expect these to order or undergo a Kondo screening only at prohibitively low temperatures}, superconducting but others are not?

\begin{figure}[tb]
\includegraphics[width=8.0cm]{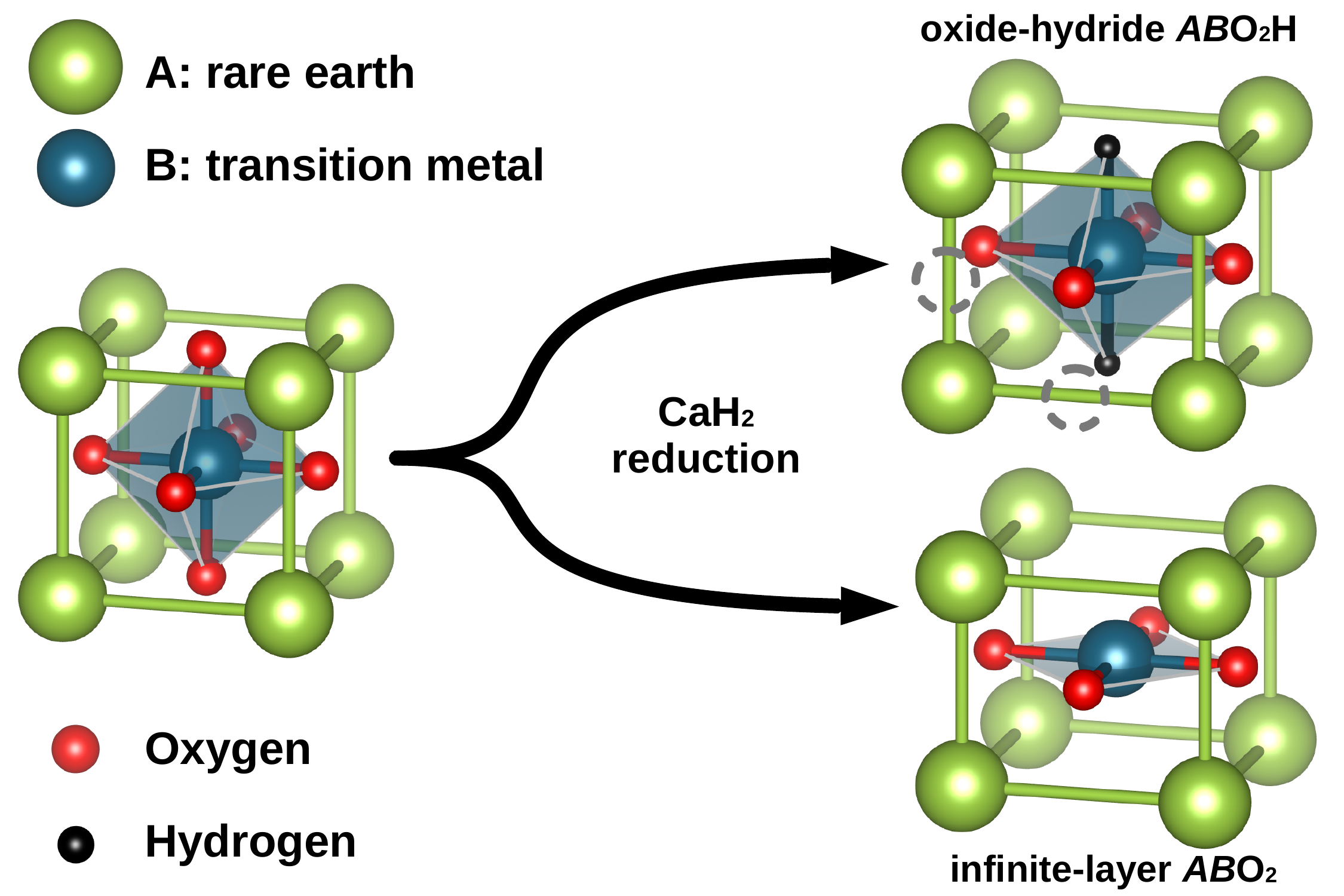}
\caption{Two possible products in the topotactic reduction of $AB$O$_3$ by means of CaH$_2$: oxide-hydride $AB$O$_2$H and infinite-layer $AB$O$_2$. The dashed circles indicate other possible H positions for $AB$O$_2$H, which are however energetically  less favorable.} 
\label{Fig1}
\end{figure}

Let us take a step back and recapitulate the synthesis of $A$NiO$_2$ nickelates with the  unusual low oxidation state Ni$^+$. It is synthesized by first growing
$A$NiO$_3$ on a SrTiO$_3$ substrate, and then reducing it to  $A$NiO$_2$ with the help of the reagent CaH$_2$,  see Fig.~\ref{Fig1}. However, there is another possible endproduct:   $A$NiO$_2$H. Indeed for another perovskite,  SrVO$_3$, it was shown in a detailed experimental analysis \cite{katayama2016epitaxial} that  the CaH$_2$ reduction reaction leads to SrVO$_2$H; also  NdNiO$_x$H$_y$ has been detected  \cite{onozuka2016formation}.

In this letter, we show based on DFT calculations  \cite{PhysRev.136.B864,PhysRev.140.A1133,Martin04} that the nickelates are just at the borderline of the two reaction paths of  Fig.~\ref{Fig1}: While  for $A$=Nd and in particular for $A$=La the oxide-hydrides NdNiO$_2$H and LaNiO$_2$H are energetically favorable, with Sr-doping, the infinite-layer Nd$_{0.8}$Sr$_{0.2}$NiO$_2$ becomes more stable.
As a matter of course, the reaction kinetics also influences the endproducts, and without carefully optimizing the reaction conditions some mixed phase of $A$NiO$_2$ and $A$NiO$_2$H may emerge.
We further demonstrate that the H intercalation  has dramatic consequences for the electronic structure as calculated by DFT and DFT+dynamical mean field theory (DMFT) \cite{RevModPhys.68.13,kotliar2004strongly,PhysRevLett.62.324,held2007electronic}:  While $A$NiO$_2$ is metallic with a very strong quasiparticle renormalization of the Ni $d_{x^2-y^2}$ band and  Nd(La)-5$d$ pocket, quite similar to doped cuprates;  $A$NiO$_2$H is a Mott insulator with two Ni bands,  $d_{x^2-y^2}$ and $d_{z^2}$, and no  Nd(La)-5$d$ pocket.


\emph{Methods.}
Structural   details and H-topotactic binding energies are computed by DFT structural relaxations  (yielding lattice constants a=b=3.889\AA, c= 3.337\AA{} for LaNiO$_2$ and   a=b=3.914\AA, c=3.383\AA{} for LaNiO$_2$H) and total energy calculations. Both \textsc{wien2k} \cite{blaha2001wien2k,schwarz02} and \textsc{vasp} \cite{PhysRevB.48.13115} codes within PBE \cite{PhysRevLett.77.3865} and PBESol \cite{PhysRevLett.100.136406} versions of the generalized gradient approximation (GGA) are employed on a $13\times13\times15$ momentum grid. The energy cut-off was  500eV in \textsc{vasp}; $R_{MT}K_{max}$=7.0 with a muffin-tin radius $R_{MT}=$2.50, 1.96, 1.69 and 1.10 a.u. for La, Ni, O and H, respectively  was employed  in \textsc{wien2k}.

For the DMFT calculations, the  \textsc{wien2k} bandstructure around the Fermi level is projected onto  Wannier functions \cite{PhysRev.52.191,RevModPhys.84.1419} using \textsc{wien2wannier} \cite{mostofi2008wannier90,kunevs2010wien2wannier} and supplemented by a local density-density interaction, taking the fully localized limit \cite{Anisimov1991} as double counting.
Since for infinite-layer LaNiO$_2$,   one  La-$d$ band crosses the Fermi level E$_f$, here a full set of La-5$d$+Ni-3$d$ bands is adopted.
For  LaNiO$_2$H, a projection onto the  Ni-3$d$ bands is possible  only  because now the La-5$d$ bands are well separated from the Ni-3$d$ bands, cf.~Supplemental Material (SM)~\cite{SM} for La-$d$ bands included in DMFT. The interaction parameters are computed by constrained random phase approximation (cRPA) \cite{PhysRevB.77.085122}:  average inter-orbital interaction $U'= 3.10\,$eV ($2.00\,$eV) and Hund's exchange  $J=0.65\,$eV  (0.25\,eV) for Ni (La). The intra-orbital Hubbard interaction follows as $U=U'+2J$. These interaction parameters are close to those of  previous studies \cite{PhysRevB.73.155112,PhysRevLett.115.236802} for 3$d$ oxides. The resulting Hamiltonian is then solved at room temperature (300\,K) in DMFT using continuous-time quantum Monte Carlo simulations in the hybridization expansions \cite{RevModPhys.83.349} implemented in \textsc{w2dynamic} \cite{PhysRevB.86.155158,wallerberger2019w2dynamics};  the maximum entropy method \cite{PhysRevB.44.6011,PhysRevB.57.10287} is employed for an analytic continuation of the spectra.


\emph{Energetics of Topotactic Hydrogen.}
Besides the two cornerstone endproducts, infinite-layer $AB$O$_2$ (e.g.,\,CaCuO$_2$, SrCuO$_2$) and oxide-hydride  $AB$O$_2$H (e.g.\,SrVO$_2$H), of Fig.~\ref{Fig1}, also intermediate products such as (Ba, Sr, Ca)TiO$_{3-x}$H$_x$ \cite{ito2017excited,yajima2012epitaxial} and NdNiO$_x$H$_y$ \cite{onozuka2016formation} have been experimentally confirmed when reducing $AB$O$_3$ with CaH$_2$ upon heating \cite{yamamoto2013hydride}\footnote{Hydrogen intercalation has also been reported when CaH$_2$ reducing  further Ruddlesden-Popper oxides \cite{Jin2019,Jin2020}}. To investigate whether it is energetically favorable to intercalate hydrogen in the topotactic reaction or not,  we compute the hydrogen binding energy 
\begin{equation}
E_{B}=E[ABO_2] +  \mu[H] - E[ABO_2H].
\label{Eq1}
\end{equation}
Here,  $E[ABO_2]$ and $E[ABO_2H]$ are the total energy of $AB$O$_2$ and $AB$O$_2$H; and  $\mu[H]=E[H_2]/2$ is the chemical potential of H.  Note that $H_2$ is a typical byproduct for the reduction with CaH$_2$ and also emerges when  CaH$_2$ is in contact with H$_2$O. Hence it can be expected to be present in the reaction.
$E_{B}$ is also the difference in binding energy for the two reaction paths of Fig.~\ref{Fig1}, i.e.,  it is energetically favorable by $E_{\rm B}$ to synthesize $AB$O$_2$H  instead of $AB$O$_2$ and H$_2$/2. Of course the reaction kinetics may change the outcome and the large entropy of 1/2 H$_2$ might  change the balance
thermodynamically in favor of ABO$_2$ \footnote{The entropy of 1/2 H$_2$ at 500 K is $S=7.5 \; 10^{-4}\,$eV/K \cite{Chase1998} corresponding to  0.38 eV. On the other hand, the reduction with CaH$_2$ might lead to a rather high $H_2$ pressure;  and ABO$_2$H has a higher entropy than  ABO$_2$. Its Mott insulating spin-1 (see below)    yields $S=k_{\rm B} \ln(3) = 1\;  10^{-4}\,$eV/K corresponding to 0.05 eV at 500K; a prospective disorder entropy of say ABO$_2$H$_{0.5}$ would yield  $S=k_{\rm B} \ln(2) =  0.6 \;  10^{-4}\,$eV/K or 0.03 eV at 500K.}. But at the very least the energetics gives us a first hint whether to  expect $AB$O$_2$H  or $AB$O$_2$.

\begin{figure*}[ht]
\includegraphics[width=\linewidth]{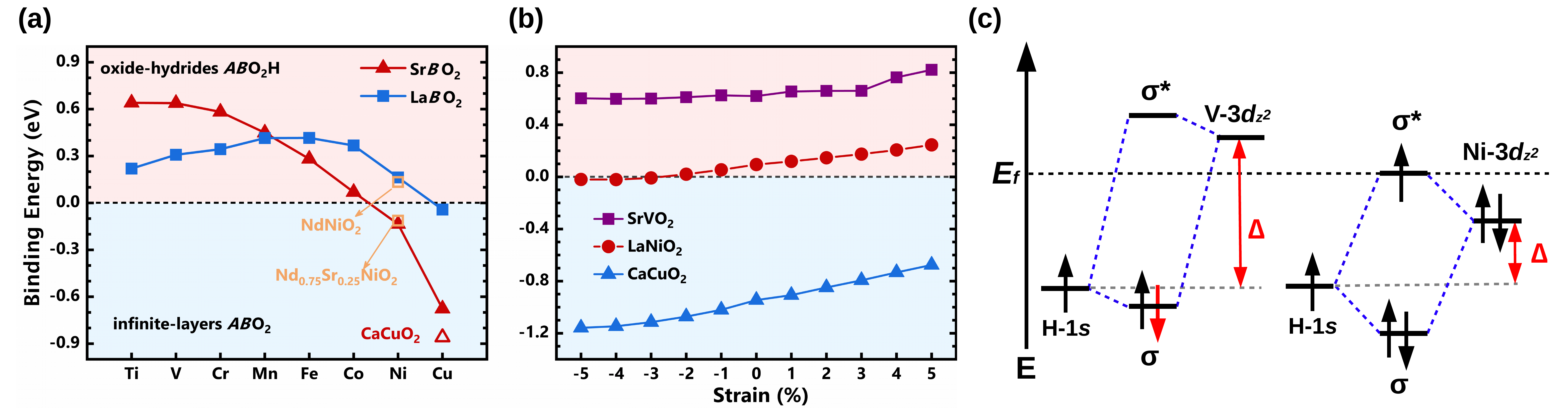}
\caption{(a) Binding energy $E_{\rm B}$ for topotactic H in  infinite-layer $AB$O$_2$ ($A$: Sr or La; $B$: Ti to Cu). Further, $E_{\rm B}$ for CaCuO$_2$, NdNiO$_2$ and Nd$_{0.75}$Sr$_{0.25}$NiO$_2$ is  plotted. (b) $E_{\rm B}$ for SrVO$_2$, LaNiO$_2$ and CaCuO$_2$  as a function of strain  at a low H-density of 12.5\%. (c)  Explanation of the evolution of the  H-topotactic binding energy from $A$VO$_2$ to  $A$NiO$_2$ based on the formation of bonding  ($\sigma$) and anti-bonding ($\sigma^*$) states between H-1$s$ and transition metal 3$d_{z^2}$. The up/down arrows indicate the electron spins, the red arrow for $A$VO$_2$ means the electron is from other (V-$t_{2g}$) orbitals.} 
\label{Fig2}
\end{figure*}

As for the three H-positions of Fig.~\ref{Fig1}, we always find that  the vacancy left by the  removed  oxygen is the energetically favored H-position. 
For a full H-topotactic intercalation, i.e.,  $AB$O$_2$H with all vacant oxygen positions occupied by H, this is plotted explicitly in Fig.~\ref{Fig1}. We first consider this complete intercalation, fully relax the $AB$O$_2$ and $AB$O$_2$H structures, and then calculate the respective total spin-unpolarized DFT energy and from this  $E_{\rm B}$ through Eq.~(\ref{Eq1}).  Our conclusions remain unchanged when using spin-polarized DFT+$U$ instead, see SM~\cite{SM}. 

For Sr$B$O$_2$ and La$B$O$_2$,  $E_{\rm B}$ is  positive from Ti to Co in Fig.~\ref{Fig2}(a), indicating the energetical preference for the  oxide-hydride Sr$B$O$_2$H for $B$=Ti$\,\ldots\,$Co. This is consistent with 
finding  SrVO$_2$H and SrTiO$_x$H$_y$ after CaH$_2$ reduction  \cite{denis2014strontium,amano2018extreme,yamamoto2017role,katayama2016epitaxial}.
Similarly, oxide-hydrides have been reported experimentally when reducing (Sr, La)CoO$_3$ \cite{hayward2002hydride,helps2010sr3co2o4} and (Ba, Ca, Sr)TiO$_3$ \cite{yajima2012epitaxial,sakaguchi2012oxyhydrides,kobayashi2012oxyhydride}. 

Surprisingly, for LaNiO$_2$, a positive $E_{\rm B}$ of 0.162\,eV is predicted, too. This indicates incorporating H topotactically in infinite-layer LaNiO$_2$ is at least energetically favorable.  
For SrNiO$_2$ on the other hand it is energetically unfavorable to intercalate H. That is, the nickelates $A$NiO$_2$ are just at the border line  $E_{\rm B}=0$ in Fig.~\ref{Fig2}(a); the cation $A$ is decisive.
The cuprates on the other hand are clearly on the  $E_{\rm B}<0$ side, i.e., hydrogen will not be intercalated,  consistent with the well studied chemistry of the cuprates.

Besides the La- and Sr-based infinite-layer $AB$O$_2$, we also calculate $E_{\rm B}$ for a few additional materials: CaCuO$_2$ \cite{siegrist1988parent} has $E_{\rm B}<0$ as the other  cuprate superconductors; undoped NdNiO$_2$ \footnote{For (Sr,\,Nd)NiO$_2$, we perform (antiferro-)magnetic GGA+$U$ calculations in \textsc{vasp} instead of the non-magnetic calculations for other $AB$O$_2$ systems because (i)  NdNiO$_3$ is experimentally determined to be an antiferromagnetic insulator \cite{li2019superconductivity} and (ii)  non-magnetic \textsc{vasp} calculations for (Sr,\,Nd)NiO$_2$ hardly converge.} has $E_{\rm B}= 0.133\,$eV, which turns negative to $E_{\rm B}=-0.113\,$eV if 25\% of Nd atoms are replaced by Sr.  Hence our  results indicate  that  infinite-layer, superconducting  Sr-doped NdNiO$_2$ is energetically stable against the topotactic inclusion of H, whereas other nickelates are not.

The complete (full) topotactic inclusion of H, where all vacancies induced by removing-oxygen are filled by H, is an ideal limiting case. Under varying experimental conditions  such as chemical reagent, substrate, temperature and strain, the H-topotactic inclusion may be incomplete, and $AB$O$_2$H$_y$ ($y$$<$1)  energetically favored. Hence we also compute $E_{\rm B}$ at a rather low H-topotactic density: $AB$O$_2$H$_{0.125}$ (achieved by including a single H into 2$\times$2$\times$2 $AB$O$_2$ supercells). Additionally, we model strain effects by changing the in-plane lattice constants ($a$, $b$), relaxing the lattice in $z$-direction and the internal atomic positions. 

Fig.~\ref{Fig2}(b) shows the corresponding $E_{\rm B}$ for SrVO$_2$H$_{0.125}$, LaNiO$_2$H$_{0.125}$, and CaCuO$_2$H$_{0.125}$.
Unstrained (0\%), the binding energy $E_{\rm B}$ per hydrogen (0.620\,eV for SrVO$_2$H$_{0.125}$, 0.094\,eV for LaNiO$_2$H$_{0.125}$ and -0.945\,eV for CaCuO$_2$H$_{0.125}$) is very similar to complete hydrogen intercalation ($E_{\rm B}=0.637\,$eV for SrVO$_2$H, 0.162\,eV for LaNiO$_2$H and -0.859\,eV for CaCuO$_2$H) in Fig.~\ref{Fig2}(a).
This indicates that the topotactic intercalation with H does not depend too much on the amount of H. 
Fig.~\ref{Fig2}(b) further shows that $E_{\rm B}$ is substantially  reduced by compressive (negative) strain. This is because the compressive strain enlarges the $z$--axis, which in turn leads to weaker  H-$B$-H bonding.
This suggests compression to be an effective way to eliminate residual H in $A$NiO$_2$. It might also explain why NdNiO$_2$ was found next to the interface of (compressive) SrTiO$_3$ , whereas NdNiO$_x$H$_y$ was found further away from the interface \cite{onozuka2016formation}.

Let us now ask: Why is $E_{\rm B}$ varying remarkably for different $AB$O$_2$? Besides the gradually changing lattice constants, the dominating factor is the $d$-band filling. By computing the band characteristics (Fig.~\ref{Fig3} below), we find H$^{-}$ which, similar to O$^{2-}_{0.5}$, absorbs one Ni electron. For early transition metals with positive $E_{\rm B}$, as e.g. SrVO$_2$H, the H-1$s$ forms bonding ($\sigma$) and anti-bonding ($\sigma^*$) states due to the orbital overlap with V-$d$$_{z^2}$
as sketched in the left panel of Fig.~\ref{Fig2}(c). The bonding $\sigma$ orbital is fully occupied with one electron originating for the H-1$s$ and the second  from the V-$t_{2g}$ orbitals. This explains the stabilization and energy gain of SrVO$_2$H.

For late transition metals, e.g. LaNiO$_2$H, the 3$d$$_{z^2}$ is fully filled, see right panel of Fig.~\ref{Fig2}(c). 
Hence, when intercalating H also the anti-bonding $\sigma^*$ orbital needs to be occupied with one electron and then crosses $E_F$.
This and the smaller bonding-antibonding splitting reduces the  energy gain for the topotactic intercalation of H.
This puts the nickelates at the borderline, whereas for the next transition metal, Cu, it is no longer energetically favorable to form $A$CuO$_2$H.

\begin{figure}[ht]
\includegraphics[width=\columnwidth]{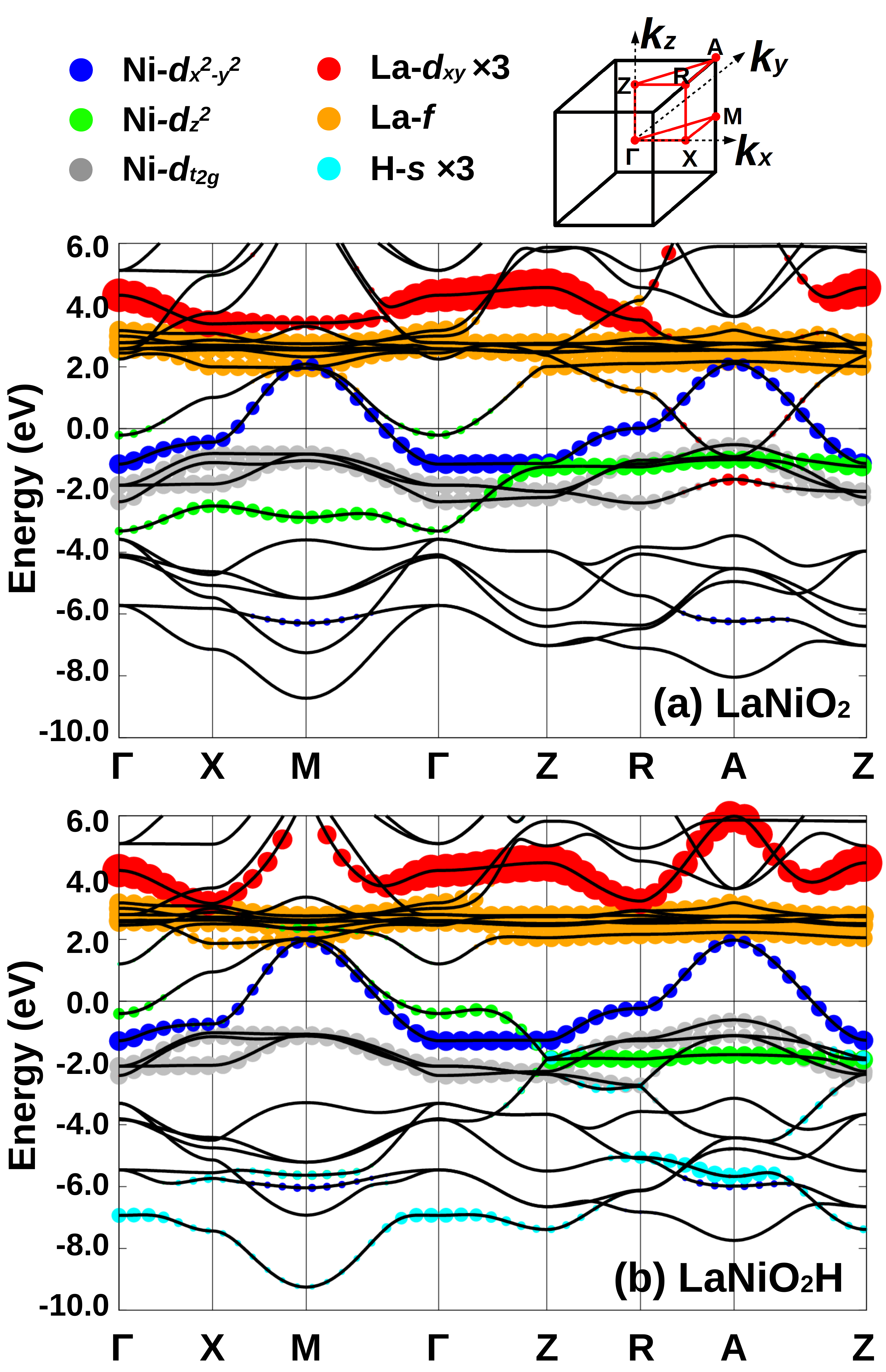}
\caption{DFT band structure and orbital characters for (a) LaNiO$_2$ and (b) LaNiO$_2$H along a high symmetry path through the Brillouin zone (top).} 
\label{Fig3}
\end{figure}
 
\emph{Spin-unpolarized DFT Electronic structure.}
Let us now address the question: How much does the electronic structure change if topotactic hydrogen is present?
In the infinite-layer LaNiO$_2$ the Ni $d_{x^2-y^2}$ orbital shows a single-band dispersion without hybridization with other bands,  similar to the Cu-$d_{x^2-y^2}$ dispersion in cuprates. However, in contrast to the cuprates there is an itinerant La-band which  crosses E$_f$ around the $A$-point, see Fig.~\ref{Fig3}(a) and \cite{Lee2004,Nomura2019,jiang2019electronic,Motoaki2019}. It is composed of of La-$5d$ but also La-$4f$ and with some Ni-$t_{2g}$ intermixing. There is no discernible hybridization gap when this 
La-band crosses the  Ni-$d_{x^2-y^2}$ band in
Fig.~\ref{Fig3}(a), because of the  (symmetry-dictated) very weak hybridization between both \cite{jiang2019electronic}.

For LaNiO$_2$H, the topotactic H  alters  the DFT band structure completely, see Fig.~\ref{Fig3}(b).
Firstly, the La band-crossing at $E_F$ around the $A$-point is gone.
This is because the  La-$d_{xy}$  hopping in the (110) direction changes sign when connected though the  H-1$s$ orbital: from $-0.098\,$eV for LaNiO$_2$ to $0.224\,$eV for LaNiO$_2$H.
This turns the minimum (La-5$d$ pocket) around the  $A$-point in
Fig.~\ref{Fig3}(a) into a maximum in Fig.~\ref{Fig3}(b).
Secondly, the Ni-$d_{z^2}$ band is now partially occupied for LaNiO$_2$H instead of being fully occupied in infinite-layer LaNiO$_2$.
Three factors contribute to this: (1) The local Ni-$d_{z^2}$ potential is shifted up by $\sim$1.5\,eV  because the $d_{z^2}$ orbitals point towards the negatively charged H$^-$. (2) The intra-orbital,  nearest-neighbor hopping of the $d_{z^2}$ electrons along $k_z$ [$\Gamma$ to Z in Fig.~\ref{Fig3} (a,b)]   changes sign  from -0.308\,eV for LaNiO$_2$  to 0.781\,eV for LaNiO$_2$H. (3) Last but not least,  H$^{-}$ reduces the valence of Ni from Ni$^{9+}$ to Ni$^{8+}$, effectively reducing  E$_f$.
Altogether, we end up in a situation which is much less akin to the cuprates:  a 3$d$$^{8}$ electronic configuration with  two Ni orbitals, 3$d_{x^2-y^2}$
and $d_{z^2}$,  but no La-5$d$ band around $E_F$. 

\begin{figure}[ht]
\includegraphics[width=\columnwidth]{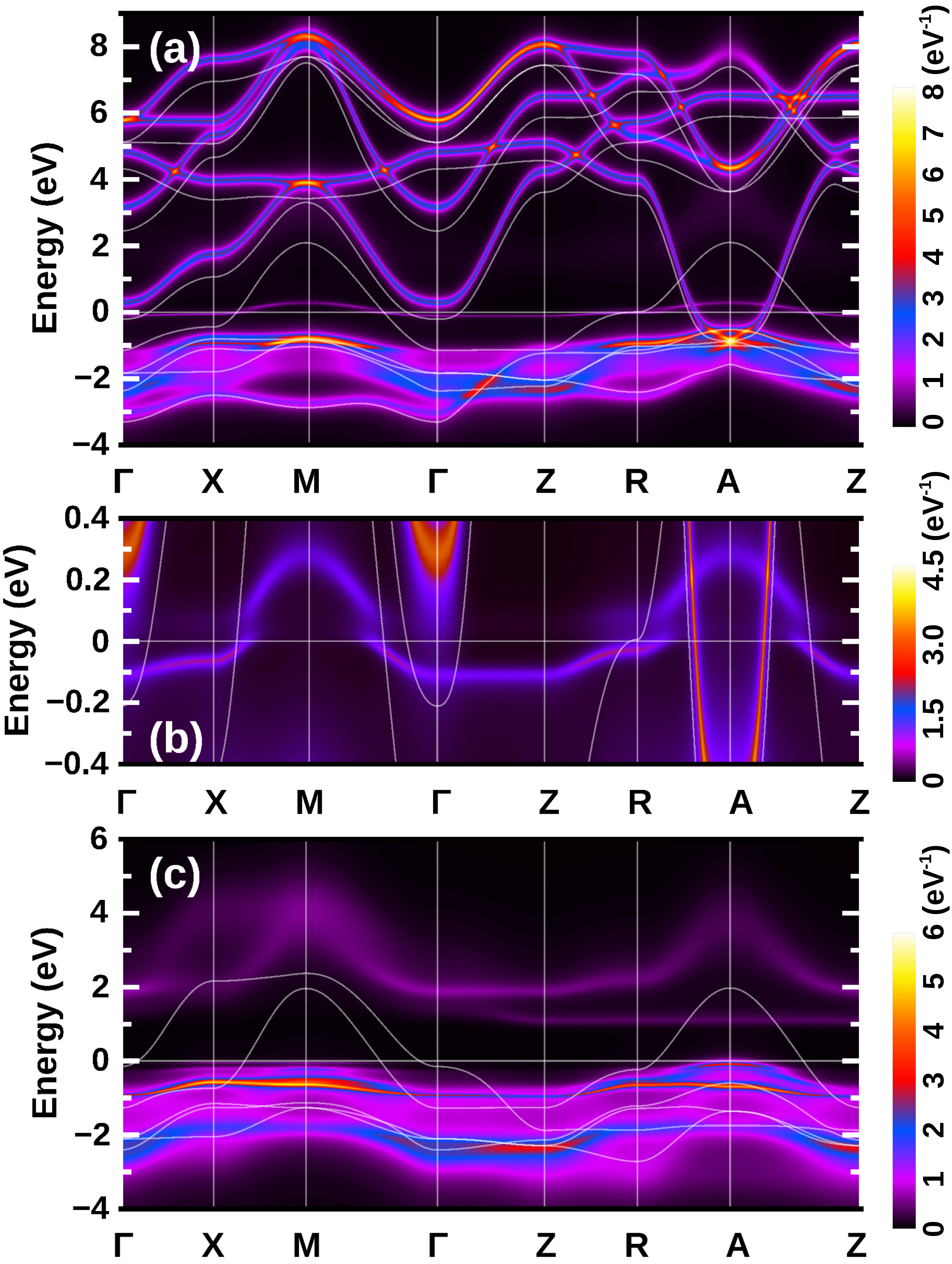}
\caption{DMFT  spectral functions $A(k,\omega)$ of (a) LaNiO$_2$  and (c) LaNiO$_2$H; solid lines: DFT Wannier-bands. Panel (b) is a zoom-in of (a) around $E_F=0$.}
\label{Fig4}
\end{figure}

\emph{DFT+DMFT electronic structure.}
In transition metal oxides, we have strong effects of electronic correlations which are not properly described by the spin-unpolarized DFT bandstructure. One can describe the opening of a Mott-like gap by  spin-polarized DFT+$U$ (see SM~\cite{SM} and \cite{PhysRevB.44.943,Mercy2017}) or by random spin orientations (or symmetry-lowering distortions) in a large supercell  \cite{Gyorffy1985,Varignon2017,Trimarchi2018,Varignon2019,Varignon2019b} \footnote{Appropriate DFT or hybrid functionals may also open bandgaps in strongly correlated systems, see e.g. \cite{PhysRevB.23.5048,Heyd2003,Tomczak2010,PhysRevLett.102.226401,Furness2018,Liu2018,Lane2018}.}. We will instead employ DFT+DMFT calculations  in the paramagnetic phase at room-temperature (300\,K) in the following. For LaNiO$_2$ [Fig.~\ref{Fig4}(a)], electronic correlations lead to a dramatic quasiparticle renormalization $Z$ or mass enhancement $m^*/m=1/Z\sim 7$ of the almost half-filled Ni 3$d_{x^2-y^2}$ band, see  zoom-in Fig.~\ref{Fig4}(b); the other Ni-3$d$ bands are almost completely filled and below $E_F$. 
 In DMFT the  La-5$d$ band still crosses $E_F$ around  the $A$-point, see Fig.~\ref{Fig4}(a), but the $\Gamma$ pocket is lifted above $E_F$ if we take the La-$d$ interaction into account (cf.\ SM~\cite{SM}). Actually its band dispersion hardly changes since  the  La-5$d$ bands are only lightly occupied.
One noteworthy  effect of electronic correlations is however to reduce the number of holes (vs.~half-filling) in the 3$d_{x^2-y^2}$ orbital, from 0.08 per Ni site for the DFT-derived Wannier Hamiltonian to 0.03 in DMFT. Without additional doping  such a light hole doping is likely not enough to induce superconductivity.  

Being so close to the Mott transition (a half-filled Ni 3$d_{x^2-y^2}$-band would be Mott insulating), minor modifications of the computational procedure such as including the La-$d$ interaction or of the experimental setup will slightly change the doping of the Ni 3$d_{x^2-y^2}$ band. And even a slight change in doping will have a big effect  on the quasiparticle renormalization (see SM~\cite{SM}). We think this explains the quite substantial variation of $Z$ in different DFT+DMFT calculations \cite{Ryee2019,Lechermann2019,Karp2020}\footnote{For a one and two band DMFT calculation cf.~\cite{Gu2019} and \cite{Werner2020}, respectively.}.

For the oxide-hydride LaNiO$_2$H, on the other hand, the La-5$d$ pockets around the $A$-point are eliminated  not only in DFT  [Fig.~\ref{Fig3}(b)] but also in DFT+DMFT [Fig.~\ref{Fig4}(c)]. Without doping through the  La-5$d$ pocket and the additional $H^-$, Ni is in an undoped 3$d^8$ configuration with holes in both the $d_{z^2}$- and $d_{x^2-y^2}$-orbitals. This integer filling of the Ni $d$-orbitals drives  LaNiO$_2$H into a Mott insulating phase  with a gap of $\sim$0.3\,eV in Fig.~\ref{Fig4}(c), a similar electronic structure for LaNiO$_2$H can also be described by   DFT+$U$, see SM~\cite{SM}. It is  consistent with the experimental observations that LaNiO$_{2.5}$, having formally the same Ni valence, was found to be  insulating \cite{PhysRevB.54.16574,abbate2002electronic}. While here we find that already the paramagnetic phase is insulating,  $C$- or $G$-type antiferromagnetic ordering \cite{Alonso1997} can be expected. Similarly SrVO$_2$H is antiferromagnetic with a high-T$_N$ \cite{denis2014strontium},  whereas SrVO$_3$ is a paramagnetic metal.

\emph{Conclusion and outlook.}
In another class of correlated superconductors, the iron pnictides \cite{Kamihara2006} \footnote{Let us note that DMFT has been quite
helpful for understanding the pnictide physics \cite{Haule2008,Aichhorn2009,Anisimov2009,Hansmann2010b,Toschi2012,Nekrasov2015,Guterding2017}}, it is well known that hydrogen plays an important role \cite{Iimura2012}, also when using CaH$_2$ as a reduction reagent \cite{Matsumoto2019}.
Here, by performing DFT and DFT+$U$ calculations, we find 
that the topotactic intercalation of hydrogen in infinite-layer $AB$O$_2$ is energetically favorable for early transition metals $B$. 

This intercalation has dramatic consequences for the electronic structures. 
LaNiO$_2$H is a 3$d$$^8$ Mott insulator with two relevant orbitals around $E_F$, 3$d_{z^2}$ and 3$d_{x^2-y^2}$, but no La-5$d$ pockets: A situation which is distinctively different from the cuprate superconductors.  Indeed this two orbital situation bears some similarities to LaNiO$_3$/LaAlO$_3$ heterostructures prior to engineering their bandstructure to a cuprate-like one \cite{Chaloupka2008,PhysRevLett.103.016401,Hansmann2010,Wang2012,Subedi2015,Janson2018}.
On the other hand, LaNiO$_2$ with a 3$d$$^9$ configuration  and the holes only in the $d_{x^2-y^2}$-orbital closely resembles the bandstructure of the doped cuprates. This  $d_{x^2-y^2}$-orbital is lightly doped already for the parent compound because of a La-5$d$ pocket around the $A$-point.

The strikingly different susceptibility toward topotactic intercalation of hydrogen and the dramatic consequences for the electronic structure may explain why some nickelates have been found to be superconducting and others not. Besides La(Nd)NiO$_2$H also an incomplete reduction to  La(Nd)NiO$_{2.5}$ \footnote{Also in \cite{Zhou2019} it has been reported to be difficult to remove the excess oxygen.} will lead to a similar Ni 3$d$$^8$ electronic configuration, and there may be mixed phases thereof in an actual experiment. While hydrogen is difficult to detect, careful studies already  confirmed its presence when reducing SrVO$_3$ \cite{katayama2016epitaxial}, SmFeAsO \cite{Matsumoto2019},  and NdNiO$_3$ \cite{onozuka2016formation,hayward1999sodium} \footnote{The hydrogen surplus ``fluorite peak'' of \cite{hayward1999sodium} has not been detected for the  superconducting nickelates in \cite{li2019superconductivity}} with the reagent CaH$_2$.

Our findings call for a careful reanalysis of the hydrogen content in nickelates. Further we suggest that for synthesizing  $A$NiO$_2$ without H and  an electronic structure prone to superconductivity,
compressive strain and Sr-doping are of advantage, as are   long reaction times for reducing the H$_2$ pressure and, according to   \cite{onozuka2016formation}, low temperatures.
Quite a number of capping layers might be needed to stabilize the system against further hydrogen intercalation, and doping with a divalent cation such as Sr appears to be necessary to arrive in the superconductive doping regime.

{\em Note added:} Further information on the synthesis used in Ref.~\onlinecite{li2019superconductivity} became recently available in Ref.~\onlinecite{Lee2020}.

\subsection{Acknowledgments}

\begin{acknowledgments}
We thank O.~Janson, P.~Hansmann, and A.~Prokofiev  for helpful comments and discussions.
L.\,S., W.\,X. and Z.\,Z. gratefully acknowledge financial support from the National Key R\&D Program of China (2017YFA0303602), 3315 Program of Ningbo, and the National Nature Science Foundation of China (11774360, 11904373). L.\,S., J.\,K., J.\,M.\,T. and K.\,H. were supported by the Austrian Science Fund (FWF) through projects P\,30997 and P\,32044. Y.\,L. acknowledges support by Deutsche Forschungsgemeinschaft (DFG) under Germany’s Excellence Strategy EXC2181/1-390900948 (the Heidelberg STRUCTURES Excellence Cluster). Calculations have been done on the Vienna Scientific Clusters (VSC) and the Supercomputing Center at NIMTE CAS.
\end{acknowledgments}

%

\end{document}